\documentclass[aps,pra,twocolumn,showpacs,amsmath,amssymb]{revtex4}

\usepackage{graphicx} 
\usepackage{dcolumn} 
\usepackage{bm} 
\usepackage{epsfig}
\usepackage{amsmath}
\usepackage{amssymb}
\usepackage{hyperref}
\usepackage{color}
\usepackage{ulem}
\begin{document}
\newcommand{\bsy}[1]{\mbox{${\boldsymbol #1}$}} 
\newcommand{\bvecsy}[1]{\mbox{$\vec{\boldsymbol #1}$}} 
\newcommand{\bvec}[1]{\mbox{$\vec{\mathbf #1}$}} 
\newcommand{\btensorsy}[1]{\mbox{$\tensor{\boldsymbol #1}$}} 
\newcommand{\btensor}[1]{\mbox{$\tensor{\mathbf #1}$}} 
\newcommand{\tensorId}{\mbox{$\tensor{\mathbb{\mathbf I}}$}} 
\newcommand{\be}{\begin{equation}}
\newcommand{\ee}{\end{equation}}
\newcommand{\bea}{\begin{eqnarray}}
\newcommand{\eea}{\end{eqnarray}}
\newcommand{\e}{\mathrm{e}}
\newcommand{\arccot}{\mathrm{arccot}}
\newcommand{\arctanh}{\mathrm{arctanh}}

\title{The electromagnetic response of a relativistic Fermi gas at finite temperatures: applications to condensed-matter systems}

\author{E. Reyes-G\'omez$^{1}$, L. E. Oliveira$^{2}$, and C. A. A. de Carvalho$^{3,4}$}
\affiliation{$^1$Instituto de F\'{i}sica, Universidad de Antioquia UdeA, Calle 70 No. 52-21, Medell\'{\i}n, Colombia \\
$^2$Instituto de F\'{i}sica, Universidade Estadual de Campinas - Unicamp, Campinas - SP, 13083-859, Brazil \\
$^3$Instituto de F\'{\i}sica, Universidade Federal do Rio de Janeiro - UFRJ, Rio de Janeiro-RJ, 21945-972, Brazil\\
$^4$Inmetro, Campus de Xer\'{e}m, Duque de Caxias-RJ, 25250-020, Brazil}

\begin{abstract}
We investigate the electromagnetic response of a relativistic Fermi gas at finite temperatures. Our theoretical results are first-order in the fine-structure constant. The electromagnetic permittivity and permeability are introduced via general constitutive relations in reciprocal space, and computed for different values of the gas density and temperature. As expected, the electric permittivity of the relativistic Fermi gas is found in good agreement with the Lindhard dielectric function in the low-temperature limit. Applications to condensed-matter physics are briefly discussed. In particular, theoretical results are in good agreement with experimental measurements of the plasmon energy in graphite and tin oxide, as functions of both the temperature and wave vector. We stress that the present electromagnetic response of a relativistic Fermi gas at finite temperatures could be of potential interest in future plasmonic and photonic investigations.
\end{abstract}


\date{\today}

\maketitle

The response of material media to applied external electromagnetic fields is central to a host of scientific and technological applications. Indeed, knowledge of how materials respond to electromagnetic perturbations, coupled to engineering at the nanoscale, has led to the construction of customized devices tailored to exhibit very specific properties. Materials obtained in this way are presently called metamaterials. Their origin goes back to a speculation by Veselago \cite{Veselago}, who investigated the then hypothetical case of media that could have negative values for both the electrical permittivity ($\varepsilon$) and the magnetic permeability ($\mu$), a behavior believed not to occur in nature. Such characteristics were, years later, obtained in artificially constructed systems made up of tiny LC-circuits, the so-called split ring resonators, responsible for enhancing magnetic responses. That was the starting point of a whole new field of research, oriented towards custom made devices, which were used to obtain perfect lenses, invisibility cloaks, and special antennae.

Recently \cite{AragaoPRDS2016}, however, it was suggested that systems with negative values for both $\varepsilon$ and $\mu$ {\sl might} occur in nature. That belief relied on the existence of candidate systems satisfying two requirements: (i) they should be  relativistic, so as to restore the symmetry between electric and magnetic effects; (ii) they should exhibit negative $\varepsilon$ for some range of frequencies, so that this behavior should naturally extend to $\mu$ for systems satisfying (i). The relativistic electron gas was then chosen as a model system that could meet both requirements, as its nonrelativistic limit was known to comply with (ii) in the long-wavelength limit. In fact, the Lindhard formula, which corresponds to the order $\alpha$ approximation to the permittivity, reduces to the Drude formula which acquires negative values for small frequencies. This behavior persists in the relativistic case, where the permeability may also acquire negative values for low frequencies \cite{AragaoPRDS2016}. Nevertheless, the fact that the model exhibits negative values for both $\varepsilon$ and $\mu$ in the long-wavelength limit at low frequencies is not enough for us to assert that this behavior will occur in nature. For that, we have to verify that the model, within the approximation used, is a reliable description of experimental results. Clearly, one needs measurements involving the relativistic electron gases to be found, for example, in synchrotron beams or in astrophysical systems for such verification. A specific experiment where an external applied field is compared to the field measured inside a synchrotron beam is definitely called for to hopefully settle the question in the near future.

The present study shows preliminary tests of the accuracy of the electron gas description by using relativistic expressions at finite temperatures, in their nonrelativistic limit, to describe quasi-free electrons in condensed matter systems, for which experimental results are already available. In particular, we have compared the dependence on temperature and wave vector of the experimental values of the electric plasmon frequencies for graphite and tin oxide wth the predictions of our formulae. Unfortunately, we cannot yet test the equivalent behavior in the magnetic case, as we are in the nonrelativistic regime.

\begin{figure}
\epsfig{file=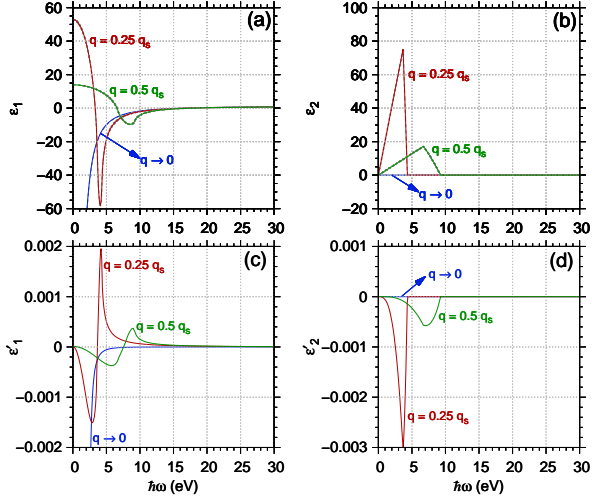,width=0.9\columnwidth}
\caption{(Color online) Real and imaginary parts of $\varepsilon$ and $\varepsilon^{\prime}$ as functions of $\hbar \omega$. Calculations were performed for various values of $q$ in units of $q_s$ (see text) and for $\eta$ corresponding to the electron density in a primitive cell of a silicon crystal. Solid and dashed lines correspond to present theoretical results [cf. Eqs. \eqref{e6} and \eqref{e7}] at $T=5$ K and to the Lindhard dielectric function \cite{Lindhard1954,WalterPRB1972}, respectively.}
\label{fig1}
\end{figure}

The constitutive relations for the electromagnetic field in a relativistic Fermi gas at finite temperature are given by \cite{AragaoPRDS2016}
\begin{subequations}
\label{e1}
\be
\label{e1a}
D^j = \varepsilon^{jk} \, E^k + \tau^{jk} \, c \, B^k,
\ee
and
\be
\label{e1b}
H^j = (\mu^{-1})^{jk} \, B^k + \sigma^{jk} \, \frac{E^k}{c}.
\ee
\end{subequations}
In the above relations, which are given in reciprocal (Fourier) space $(\bvec{q},\omega)$, one has
\be
\label{e2}
\varepsilon^{jk} = \varepsilon \, \delta^{jk} + \varepsilon^{\prime} \,\hat{q}^j \hat{q}^k,
\ee
\be
\label{e3}
(\mu^{-1})^{jk} = \mu^{-1} \, \delta^{jk} + {\mu^{\prime}}^{-1} \,\hat{q}^j \hat{q}^k,
\ee
\be
\label{e4}
\tau^{jk} = \tau \, \epsilon^{jkl} \, \hat{q}^l,
\ee
and
\be
\label{e5}
\sigma^{jk} = \sigma \, \epsilon^{jkl} \, \hat{q}^l,
\ee
where $\delta^{jk}$ is the Kronecker delta, $\epsilon^{jkl}$ is the Levi-Civita symbol, $\hat{q}^j = q^j / q$, $q = \vert \bvec{q} \vert$,
\be
\label{e6}
\varepsilon = 1 + {\cal A} + \left ( 1 - \frac{\tilde{\omega}^2}{\tilde{q}^2} \right ) {\cal B} + \left ( 2 - \frac{\tilde{\omega}^2}{\tilde{q}^2 - \tilde{\omega}^2} \right ) {\cal C},
\ee
\be
\label{e7}
\mu^{-1} = 1 + {\cal A} - 2 \frac{\tilde{\omega}^2}{\tilde{q}^2} {\cal B} + \left ( 2 + \frac{\tilde{q}^2}{\tilde{q}^2 - \tilde{\omega}^2} \right ) {\cal C},
\ee
\be
\label{e8}
\varepsilon^{\prime} = - {\mu^{\prime}}^{-1} = - {\cal A} + \frac{\tilde{q}^2}{\tilde{q}^2 - \tilde{\omega}^2} {\cal C},
\ee
and
\be
\label{e9}
\tau = \sigma = \frac{\tilde{\omega}}{\tilde{q}} \left [ - {\cal B} + \frac{\tilde{q}^2}{\tilde{q}^2 - \tilde{\omega}^2} {\cal C} \right ].
\ee

Here we have defined the dimensionless variables $\tilde{q} = q / q_c$, $\tilde{\omega} = \omega / \omega_c$, $\tilde{\beta} = m c^2 \beta$, and $\tilde{\xi} = \xi/ (m c^2)$, where  $q_c = m c / \hbar$ is the Compton wave vector, $\omega_c = m c^2 / \hbar$ is the Compton frequency, $\beta = 1 / (k_B T)$, $T$ is the absolute temperature, and $\xi$ is the chemical potential of the Fermi gas. The dimensionless scalar functions in Eqs. \eqref{e6}-\eqref{e9} are given by
\be
\label{e10}
{\cal A} =  \frac{4 \alpha}{\pi} \frac{1}{\tilde{q}^2-\tilde{\omega}^2} \left [ {\cal I}  + \left ( 1 - \frac{3}{2} \frac{\tilde{q}^2-\tilde{\omega}^2}{\tilde{q}^2} \right )  {\cal J} \right ],
\ee
\be
\label{e11}
{\cal B} =  \frac{4 \alpha}{\pi} \frac{{\cal J}}{\tilde{q}^2-\tilde{\omega}^2},
\ee
and
\be
\label{e12}
{\cal C} = - \frac{\alpha}{3 \pi} \biggl \{ \frac{1}{3} + \left ( 3 + \gamma^2 \right )  \left [ \gamma \, \arccot (\gamma)  -1  \right ] \biggr \},
\ee
where $\alpha$ is the fine-structure constant,
\be
\label{e13}
{\cal I}  =  \int_{0}^{\infty}  dy \, \frac{y^2}{\sqrt{y^2 + 1}} {\cal F}_0  \left [ 1  +  \frac{2  -  \tilde{q}^2  +  \tilde{\omega}^2}{8 y \tilde{q}} {\cal F}_1 \right ],
\ee
\bea
\label{e14}
{\cal J} &=& \int_{0}^{\infty} dy \, \frac{y^2}{\sqrt{y^2+1}} {\cal F}_0 \left [ 1 + \frac{4 (y^2+1) - \tilde{q}^2 + \tilde{\omega}^2}{8 y \tilde{q}} {\cal F}_1 \right. \nonumber \\ &-& \left. \frac{\tilde{\omega} \sqrt{y^2+1}}{2 y \tilde{q}} {\cal F}_2  \right ],
\eea
\be
\label{e15}
{\cal F}_0 \! \left ( y,\tilde{\beta},\tilde{\xi} \right ) \! = \! \frac{1}{\e^{\tilde{\beta} \left ( \sqrt{y^2 + 1} - \tilde{\xi} \right )} \! + \! 1} \! - \! \frac{1}{\e^{\tilde{\beta} \left ( \sqrt{y^2 + 1} + \tilde{\xi} \right )} \! + \! 1},
\ee
\be
\label{e16}
{\cal F}_1 (y,\tilde{q},\tilde{\omega}) \! = \! \ln \! \left [ \frac{(\tilde{q}^2 \! - \! \tilde{\omega}^2 + 2 y \tilde{q})^2 \! - \! 4 (y^2 + 1) \tilde{\omega}^2}{(\tilde{q}^2 \! - \! \tilde{\omega}^2 - 2 y \tilde{q})^2 \! - \! 4 (y^2 + 1) \tilde{\omega}^2} \right ] \!,
\ee
\be
\label{e17}
{\cal F}_2 (y,\tilde{q},\tilde{\omega}) \! = \! \ln \! \left [ \! \frac{\tilde{\omega}^4 - 4 (\tilde{\omega} \sqrt{y^2+1} + y \tilde{q})^2}{\tilde{\omega}^4 - 4 (\tilde{\omega} \sqrt{y^2+1} - y \tilde{q})^2} \! \right ] \!,
\ee
and
\be
\label{e18}
\gamma = \sqrt{\frac{4}{\tilde{q}^2-\tilde{\omega}^2}-1}.
\ee

The electromagnetic response of a relativistic Fermi gas at finite temperature may be straightforwardly evaluated through Eqs. \eqref{e6}-\eqref{e9} by computing the integrals given by Eqs. \eqref{e13} and \eqref{e14}. To do that, it is first necessary to obtain the $\xi$ chemical potential of the  relativistic Fermi gas by solving \cite{AragaoPRDS2016} the transcendental equation 
\be
\label{e19}
\Delta N = N^- - N^+ = \sum_{\bvec{p}} g \, f_0 (p,\beta,\xi),
\ee
where $N^-$ and $N^+$ are the number of particles and antiparticles in the Fermi gas, respectively, 
\be
\label{e20}
f_0 (p,\beta,\xi) = \frac{1}{\e^{\beta \left ( \Omega_p - \xi \right )} + 1} - \frac{1}{\e^{\beta \left ( \Omega_p + \xi \right )} + 1}
\ee
is the distribution function accounting for the presence of both particles and antiparticles, $\Omega_p = \sqrt{p^2 c^2 + m^2 c^4}$ is the relativistic energy of a carrier with momentum $p$, and $g=2$ is the spin degeneracy factor of the Fermi gas. By defining the effective carrier density $\eta = \Delta N/V$, Eq. \eqref{e19} reduces to
\be
\label{e22}
\eta = \eta_0 \int_{0}^{+ \infty} dy \, y^2 \, {\cal F}_0 \left ( y,\tilde{\beta},\tilde{\xi} \right )
\ee
where $\eta_0 = g m^3 c^3 / (2 \pi^2 \hbar^3) \approx 1.76 \times 10^{30}$ cm$^{-3}$. 

\begin{figure}
\epsfig{file=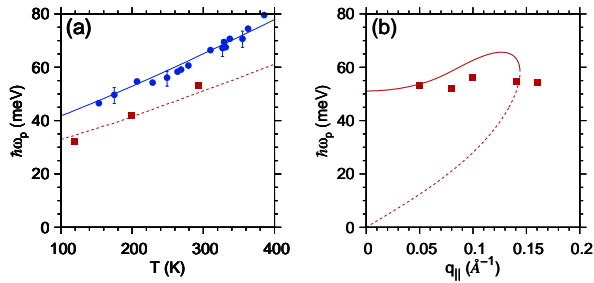,width=0.9\columnwidth}
\caption{(Color online) (a) Temperature dependence of the plasmon energy of graphite. Dots and squares correspond to the experimental values by Jensen {\it et al} \cite{JensenPRL1991} and Portail {\it et al} \cite{PortailSS1999}, respectively, for plasmons excited by an electromagnetic field with electric-field component parallel to the unitary vector $\bvec{c}$ normal to the basal plane. Solid and dashed lines correspond to theoretical results obtained from Eq. \eqref{e6}, in the limit $q \rightarrow 0$, by replacing the free-electron mass by the effective masses $m_c^* = 3.7 \, m$ and $m_c^* = 6 \, m$, respectively. The $\eta$ carrier density was assumed as a temperature-dependent function (see text). (b) Plasmon energy of graphite as a function of the $q_{\parallel}$ wave vector parallel to the $\bvec{c}$ direction. Solid squares correspond to the experimental measurements by Portail {\it et al} \cite{PortailSS1999} at $T=300$ K, whereas the solid line corresponds to present calculations from Eq. \eqref{e6}, at the same value of $T$, computed by taking \cite{ChungJMS2002} $m_c^* = 6 \, m$.  The dashed line corresponds to the lower zero of Eq. \eqref{e6}. The $\eta$ carrier density was taken as in panel (a).}
\label{fig2}
\end{figure}

\begin{figure}
\epsfig{file=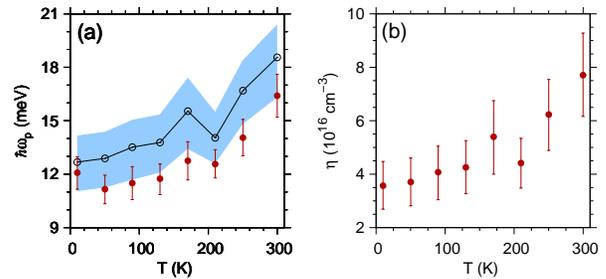,width=0.9\columnwidth}
\caption{(Color online) (a) Plasmon energy of tin oxide nanowire films as a function of the $T$ temperature. Solid circles correspond to experimental measurements by Zou {\it et al} \cite{ZouJPDAP2012}. Open circles correspond to numerical results obtained from Eq. \eqref{e6} by using the carrier density, at each value of $T$, reported by Zou {\it et al} \cite{ZouJPDAP2012} [see panel (b)]. Calculations were performed  in the limit $q \rightarrow 0$ by replacing the free-electron mass by the  $m_c^* = 0.31 \, m$ conduction-effective mass of tin oxide \cite{ZouJPDAP2012}. The solid line connecting open circles is a guide to the eye. The dark area corresponds to the uncertainty interval of the calculated plasmon energy at each value of $T$ and was computed by propagating the error of the carrier density estimated by the error bars in panel (b).}
\label{fig3}
\end{figure}

We denote $\varepsilon_i$ and $\varepsilon^{\prime}_i$ as the real ($i=1$) and imaginary ($i=2$) parts of the electromagnetic-response functions given by Eqs. \eqref{e6} and \eqref{e7}, respectively. First we have focused on the dielectric tensor defined by Eq. \eqref{e2}. It is possible to see that $\left ( \varepsilon^{jk} \right )$ may be diagonalized $\mathrm{diag} (\varepsilon,\varepsilon,\varepsilon + \varepsilon^{\prime})$. We display in Fig. \ref{fig1} the real and imaginary parts of $\varepsilon$ and $\varepsilon^{\prime}$ as functions of $\hbar \omega$. Results were obtained for $\eta$ corresponding to the electron density in a primitive cell of a silicon crystal and for various values of the wave vector $q$ expressed in units of $q_s = 2 \pi / a$. Solid lines correspond to present numerical results obtained at $T=5$ K, whereas dashed lines correspond to the Lindhard dielectric function \cite{Lindhard1954,WalterPRB1972}. As expected, results computed from Eq. \eqref{e6} coincide with the Lindhard dielectric function in the non-relativistic limit. It is apparent from Fig. \ref{fig1} that the electric permittivity is essentially isotropic, in the non-relativistic limit, due to the negligible contribution of $\varepsilon^{\prime}$ to the $\left ( \varepsilon^{jk} \right )$ tensor [cf. Figs. \ref{fig1}(c) and \ref{fig1}(d)].

Present theoretical results may be used to compute the plasma frequency (or equivalently, the plasmon energy) as function of the system temperature. It is well known that the plasma frequency $\omega_p$ corresponds to the upper frequency zero of the real part of the electric permittivity \cite{WalterPRB1972}. In the non-relativistic limit, the zeroes of the eigenvalues of the dielectric tensor essentially coincide due to its almost-isotropic behavior. Therefore, one may compute the plasma frequency of the Fermi gas from one of the eigenvalues of $\left ( \varepsilon^{jk} \right )$. Here, we have compared present theoretical results for the plasmon energy $\hbar \omega_p$ with some experimental measurements. In this respect, the experimental dependencies of the plasmon energy of graphite as a function of the $T$ temperature, reported by Jensen {\it et al} \cite{JensenPRL1991} and Portail {\it et al} \cite{PortailSS1999}, are depicted in Fig. \ref{fig2}(a). In all cases, the plasmon modes were excited by an electromagnetic field with electric-field component parallel to the unitary vector $\bvec{c}$. Solid and dashed lines correspond to present theoretical results obtained from Eq. \eqref{e6}, where we have replaced the free-electron mass by the estimated effective masses $m_c^* = 3.7 \, m$ and $m_c^* = 6 \, m$, respectively, in the $\bvec{c}$ direction normal to the basal plane. Calculations were performed in the limit $q \rightarrow 0$. Here, we have considered the $\eta$ carrier density as a function of the temperature and performed a parabolic fitting in $T$ of the experimental data \cite{SpainPTRSL1967,ChungJMS2002} $\eta = 3 \times 10^{18}$ cm$^{-3}$ at $T=4.2$ K, $\eta = 4.2 \times 10^{18}$ cm$^{-3}$ at $T=77.5$ K, and $\eta = 1.13 \times 10^{19}$ cm$^{-3}$ at $T=300$ K. The use of the effective mass $m_c^* = 3.7 \, m$ leads to a good agreement between present theoretical calculations and experimental results by Jensen {\it et al} \cite{JensenPRL1991}.  It should be noted that this value of the effective mass is smaller that the effective mass $m_c^* = 6 \, m$ reported by  Chung \cite{ChungJMS2002}, which leads to a good agreement between present results and the experimental measurements performed by Portail {\it et al} \cite{PortailSS1999}. Of course, an appropriate study of the plasmon energy as a function of the temperature should include the band structure information of the specific material. Obtaining the plasmon energy from the dielectric function of an electron gas necessarily implies the use of fitting parameters, such as the effective mass and the carrier density, in order to describe the experimental results. Moreover, even in a quantum-mechanical calculation, the specific conditions of each experiment should be considered in detail to account for the strong dispersion observed in the experimental results [see, for instance, the remarkable differences between the experimental results by Jensen {\it et al} \cite{JensenPRL1991} and Portail {\it et al} \cite{PortailSS1999} depicted in Fig. \ref{fig2}(a)]. The behavior of the plasmon energy of graphite, as a function of the $q_{\parallel}$ wave vector parallel to the $\bvec{c}$ direction, is depicted in Fig. \ref{fig2}(b). Solid symbols correspond to the experimental measurements reported by Portail {\it et al} \cite{PortailSS1999} at $T=300$ K, whereas the solid line corresponds to present theoretical results obtained from Eq. \eqref{e6} at this temperature value. In addition, we have shown in Fig. \ref{fig2}(b) the behavior of the lowest-frequency zero of $\varepsilon_1$ [cf. dashed line in Fig. \ref{fig2}(b)], given by Eq. \eqref{e6}, as a function of $T$. The free-electron mass was replaced by $m_c^* = 6 \, m$ for computing purposes \cite{ChungJMS2002}. Moreover, the $\eta$ carrier density was taken as in Fig. \ref{fig2}(a). Once again, present theoretical results are consistent with the experimental measurements by Portail {\it et al} \cite{PortailSS1999}. The overall behavior of the dielectric function as a function of the wave vector is quite similar to that obtained by Yi and Kim, in wurtzite GaN, from non-relativistic RPA theoretical calculations \cite{YiPB2015}.

Finally, the temperature dependence of the plasmon energy of tin oxide nanowire films is displayed in Fig. \ref{fig3}(a). Solid and open circles correspond to measurements reported by Zou {\it et al} \cite{ZouJPDAP2012} and theoretical calculations from Eq. \eqref{e6}  in the limit $q \rightarrow 0$, respectively.  Present theoretical values of the plasmon energy were obtained by using the carrier densities reported by Zou {\it et al} \cite{ZouJPDAP2012} at specific values of $T$, as depicted in Fig. \ref{fig3}(b). The dark area in Fig. \ref{fig3}(a) corresponds to the uncertainty interval corresponding to the calculated plasmon energy, which was computed by propagating the error of the carrier density estimated by the error bars in Fig. \ref{fig3}(b). Here, we have replaced the free-electron mass by the  $m_c^* = 0.31 \, m$ conduction-effective mass of tin oxide \cite{ZouJPDAP2012}. In this case, present theoretical calculations slightly overestimate the plasmon energy as compared with the experimental results. Nevertheless, the tolerance intervals of experimental and theoretical data overlap, which indicates good agreement between both results.

Summing up, in the present work we have investigated the electromagnetic response of a relativistic Fermi gas at finite temperatures, in the nonrelativistic regime. Generalized permittivities and permeabilities, which result from a first-order correction on the fine-structure constant, were introduced through general constitutive relations in reciprocal space. Results obtained in the limits of low temperatures and low carrier densities were used to study the behavior of the electric plasmon energy, as a function of the temperature and wave vector, in some condensed-matter systems such as graphite and tin oxide. The plasmon energy was calculated from the electric permittivity and found in good agreement with previous experimental measurements in such systems. We do hope that present theoretical results will be of importance in condensed-matter applications involving plasmonics and photonics.

\acknowledgments

The authors would like to thank the Scientific Colombian Agency CODI - University of Antioquia, and Brazilian Agencies CNPq, FAPESP (Procs. 2012/51691-0 and 2013/21320-3), and FAEPEX-UNICAMP for partial financial support.

\end{document}